\documentclass[cup6b]{cupbook}
\usepackage{natbib}
\usepackage{natpatch}
\usepackage{epsf}
\usepackage{graphicx}

\newcommand\lsim{\mathrel{\rlap{\lower4pt\hbox{\hskip1pt$\sim$}}
    \raise1pt\hbox{$<$}}}
\newcommand\gsim{\mathrel{\rlap{\lower4pt\hbox{\hskip1pt$\sim$}}
    \raise1pt\hbox{$>$}}}
\def\lsim{\mathrel{\raise.3ex\hbox{$<$\kern-.75em\lower1ex\hbox{$\sim$}}}} 
\def\gsim{\mathrel{\raise.3ex\hbox{$>$\kern-.75em\lower1ex\hbox{$\sim$}}}}

\newcommand{\tev}{\,\mbox{TeV}}
\newcommand{\gev}{\,\mbox{GeV}}
\newcommand{\mev}{\,\mbox{MeV}}
\newcommand{\kev}{\,\mbox{keV}}

\newcommand{\pb}{\,\mbox{pb}}

\newcommand{\text}[1]{{\rm #1}}
\newcommand{\mweak}{m_{\rm weak}}    
\newcommand{\gweak}{g_{\rm weak}}    
\newcommand{\mgravitino}{m_{\gravitino}}
\newcommand{\gravitino}{\tilde{G}}
\newcommand{\NLSP}{\text{NLSP}}

\newcommand{\mstar}{M_{*}}

\newcommand{\eqref}[1]{Eq.~(\ref{#1})}
\newcommand{\figref}[1]{Fig.~\ref{fig:#1}}
\newcommand{\s}{\text{s}}

\newcommand{\WIMP}{\text{WIMP}}
\newcommand{\SWIMP}{\text{SWIMP}}
\newcommand{\mWIMP}{m_{\WIMP}}
\newcommand{\mSWIMP}{m_{\SWIMP}}
\newcommand{\YWIMP}{Y_{\WIMP}}

\newcommand{\Bino}{\tilde{B}}

\newcommand{\stau}{\tilde{\tau}}

\newcommand{\epsEM}{\varepsilon_{\text{EM}}}

\newcommand{\zetaEM}{\zeta_{\text{EM}}}

\newcommand{\ThBBN}{T^h_{\text{BBN}}}

\newcommand{\TRH}{T_{\text{RH}}}

\newcommand{\ThRH}{T^h_{\text{RH}}}

\newcommand{\ghlightBBN}{g^{h\, \text{BBN}}_{\text{light}}}
\newcommand{\ghheavyBBN}{g^{h\, \text{BBN}}_{\text{heavy}}}

\newcommand{\sigmaSI}{\sigma_{\text{SI}}}
\newcommand{\rem}[1]{{}}

\begin{document}

\cleardoublepage
\pagenumbering{arabic}


\author[J.~L.~Feng]{Jonathan L.~Feng \\ Department of Physics and
  Astronomy, University of California, Irvine, CA 92697, USA}

\setcounter{chapter}{9}
\chapter[Non-WIMP Candidates]{Non-WIMP Candidates$^\dagger$}

\section{Motivations}

\footnote[0]{\hspace*{-.07in}$^\dagger$Published as Chapter 10,
pp.~190-204, in {\em Particle Dark Matter: Observations, Models and
Searches}, edited by Gianfranco Bertone (Cambridge University Press,
2010), available at
http://cambridge.org/us/catalogue/catalogue.asp?isbn=9780521763684.}There
are many non-WIMP dark matter candidates.  Two prominent and highly
motivated examples are axions and sterile neutrinos, which are
reviewed in Chapters~11\rem{\ref{Chap:Sikivie}} and
12\rem{\ref{Chap:Shaposhnikov}}, respectively.  In addition, there are
candidates motivated by minimality, particles motivated by
experimental anomalies, and exotic possibilities motivated primarily
by the desire of their inventors to highlight how truly ignorant we
are about the nature of dark matter.

In this brief Chapter, we focus on dark matter candidates that are not
WIMPs, but which nevertheless share the most important virtues of
WIMPs.  As discussed in Chapters~7\rem{\ref{Chap:Gelmini}},
8\rem{\ref{Chap:Servant}}, and 9\rem{\ref{Chap:Ellis}}, WIMPs have
several nice properties:
\begin{itemize}
\item They exist in well-motivated particle theories.  
\item They are naturally produced with the correct thermal relic
  density (the ``WIMP miracle'').
\item They predict signals that may be seen in current and near future
  experiments.
\end{itemize}
The candidates we discuss also have all three of these properties.
They fall naturally into two classes: superWIMP candidates, which
inherit the correct relic density through decays, and WIMPless
candidates, which have neither weak-scale masses nor weak
interactions, but which nevertheless have the correct thermal relic
density.  These possibilities appear in the same particle physics
frameworks as WIMPs, but they imply very different cosmological
histories for our Universe, as well as qualitatively new dark matter
signals for both astrophysical observatories and particle physics
experiments.

\section{SuperWIMP Dark Matter}

\subsection{Candidates and Relic Densities}

\begin{figure*}[tb]
\centering
\includegraphics[height=10cm,angle=-90]{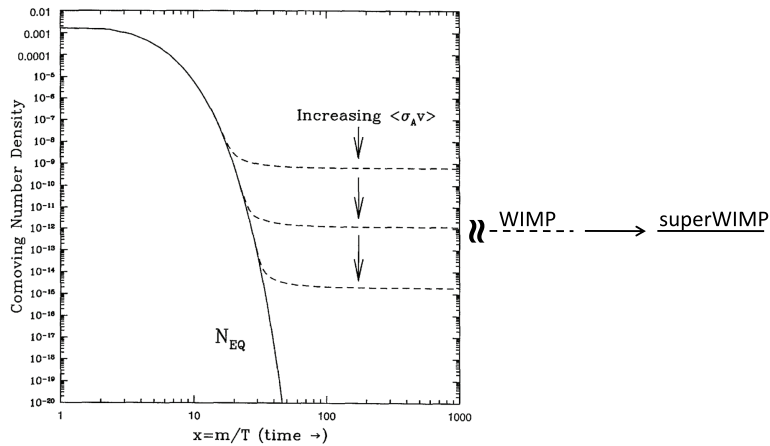}
\caption{In superWIMP scenarios, WIMPs freeze out as usual, but then
decay to superWIMPs, superweakly-interacting massive particles that
form dark matter.}
\label{fig:freezeout_swimp}
\end{figure*}

In the superWIMP framework for dark matter, WIMPs freeze out as usual
in the early Universe, but later decay to superWIMPs,
superweakly-interacting massive particles that form the dark matter
that exists today.  Because superWIMPs are very weakly-interacting,
they have no impact on WIMP freeze out in the early universe, and the
WIMPs decouple, as usual, with a thermal relic density
$\Omega_{\text{WIMP}}$ that is naturally near the required density
$\Omega_{\text{DM}} \approx 0.23$.  Assuming that each WIMP decay
produces one superWIMP, the relic density of superWIMPs is
\begin{equation}
\Omega_{\text{SWIMP}} = \frac{m_{\text{SWIMP}}}{m_{\text{WIMP}}}
\Omega_{\text{WIMP}} \ .
\end{equation}
SuperWIMPs therefore inherit their relic density from WIMPs, and for
$m_{\text{SWIMP}} \sim m_{\text{WIMP}}$, they are also naturally
produced in the desired amount to be much or all of dark matter.  The
evolution of number densities is shown in \figref{freezeout_swimp}.

The superWIMP scenario is realized in many particle physics models.
The prototypical example is gravitinos, which exist in all
supersymmetric theories~\cite{Feng:2003xh,Feng:2003uy,%
Feng:2003nr,Ellis:2003dn,Buchmuller:2004rq,Wang:2004ib,Feng:2004zu,%
Feng:2004mt,Feng:2004gn,Ellis:2004bx,Roszkowski:2004jd}.  In the
simplest supersymmetric models, supersymmetry is transmitted to
standard model superpartners through gravitational interactions, and
supersymmetry is broken at a high scale.  The mass of the gravitino
$\gravitino$ is
\begin{equation}
m_{\gravitino} = \frac{F}{\sqrt{3} \mstar} \ ,
\label{gravitinomass}
\end{equation}
and the masses of standard model superpartners are
\begin{equation}
\tilde{m} \sim \frac{F}{\mstar} \ ,
\label{tildem}
\end{equation}
where $\mstar = (8 \pi G_N)^{-1/2} \simeq 2.4 \times 10^{18}~\gev$ is
the reduced Planck scale and $F \sim (10^{11}~\gev)^2$ is the
supersymmetry-breaking scale squared.  The precise ordering of masses
depends on unknown, presumably ${\cal O}(1)$, constants in
\eqref{tildem}. It is, then, perfectly possible that the gravitino is
the lightest supersymmetric particle (LSP) and a candidate for
superWIMP dark matter. The role of the decaying WIMP is played by the
next-to-lightest supersymmetric particle (NLSP), typically a slepton,
sneutrino, or neutralino.  As required, the gravitino couples very
weakly, with interactions suppressed by $\mstar$, and so it is
irrelevant during the WIMP's thermal freeze out.

The gravitino superWIMP scenario differs markedly from other gravitino
dark matter scenarios.  In previous
frameworks~\cite{Pagels:1981ke,Weinberg:1982zq,Krauss:1983ik,%
Nanopoulos:1983up,Khlopov:1984pf,Ellis:1984eq,Ellis:1984er,%
Juszkiewicz:1985gg,Ellis:1990nb,Moroi:1993mb,Bolz:2000fu}, gravitinos
were expected to be produced either thermally, with
$\Omega_{\gravitino} \sim 0.1$ obtained by requiring $\mgravitino \sim
\kev$, or through reheating, with $\Omega_{\gravitino} \sim 0.1$
obtained by tuning the reheat temperature to $T_{\text{RH}} \sim
10^{10}~\gev$.  In the superWIMP scenario, the desired amount of dark
matter is obtained without relying on the introduction of new,
fine-tuned energy scales.

Other examples of superWIMPs include Kaluza-Klein gravitons in
scenarios with universal extra
dimensions~\cite{Feng:2003xh,Feng:2003uy,Feng:2003nr},
axinos~\cite{Rajagopal:1990yx,Covi:1999ty,Covi:2001nw} and
quintessinos~\cite{Bi:2003qa,Bi:2004ys} in supersymmetric theories,
and many other scenarios in which a metastable particle decays to the
true dark matter particle through highly suppressed interactions, with
lifetimes ranging from fractions of a second to beyond the age of the
Universe.

\subsection{Astrophysical Signals}

Because superWIMPs are very weakly interacting, they are impossible to
detect in conventional direct and indirect dark matter search
experiments.  At the same time, the extraordinarily weak couplings of
superWIMPs imply that the decays of WIMPs to superWIMPs may be very
late and have an observable impact on, for example, Big Bang
nucleosynthesis (BBN), the Planckian spectrum of the cosmic microwave
background (CMB), small scale structure, the diffuse photon flux, and
cosmic ray experiments.

In the prototypical case of a slepton decaying to a gravitino
superWIMP, the decay width is
\begin{equation}
 \Gamma(\tilde{l} \to l \tilde{G}) =\frac{1}{48\pi \mstar^2}
 \frac{m_{\tilde{l}}^5}{m_{\tilde{G}}^2} 
 \left[1 -\frac{m_{\tilde{G}}^2}{m_{\tilde{l}}^2} \right]^4 \ ,
\label{sfermionwidth}
\end{equation}
assuming the lepton mass is negligible. This decay width depends on
only the slepton mass, the gravitino mass, and the Planck mass.  For
$\mgravitino / m_{\tilde{l}} \approx 1$, the slepton decay lifetime is
\begin{eqnarray}
 \tau(\tilde{l} \to l \tilde{G})
\simeq 3.6\times 10^8~\s
\left[\frac{100~\gev}{m_{\tilde{l}} - m_{\gravitino}}\right]^4
\left[\frac{m_{\tilde{G}}}{\tev}\right]\ .
\label{decaylifetime}
\end{eqnarray}
This expression is valid only when the gravitino and slepton are
nearly degenerate, but usefully illustrates that decay lifetimes of
the order of days or months are perfectly natural. Similar expressions
hold for the decay of a neutralino NLSP to a gravitino.

\subsubsection{BBN and CMB}

Signals in BBN and the CMB are determined primarily by the WIMP
lifetime and the energy released in visible decay products when the
WIMP decays.  This energy release destroys and creates light elements,
distorting the predictions of standard BBN.  In addition, the
injection of electromagnetic energy may also distort the frequency
dependence of the CMB away from its ideal black body spectrum.  For
the decay times of interest with redshifts $z \sim 10^5$ to $10^7$,
the resulting photons interact efficiently through $\gamma e^- \to
\gamma e^-$ and $e X \to e X \gamma$, where $X$ is an ion, but photon
number is conserved, since double Compton scattering $\gamma e^- \to
\gamma \gamma e^-$ is inefficient.  The spectrum therefore relaxes to
statistical but not thermodynamic equilibrium, resulting in a
Bose-Einstein distribution function
\begin{equation}
f_{\gamma}(E) = \frac{1}{e^{E/(kT) + \mu} - 1} \ ,
\end{equation}
with chemical potential $\mu \ne 0$.

\begin{figure}[tb]
\includegraphics*[width=10cm]{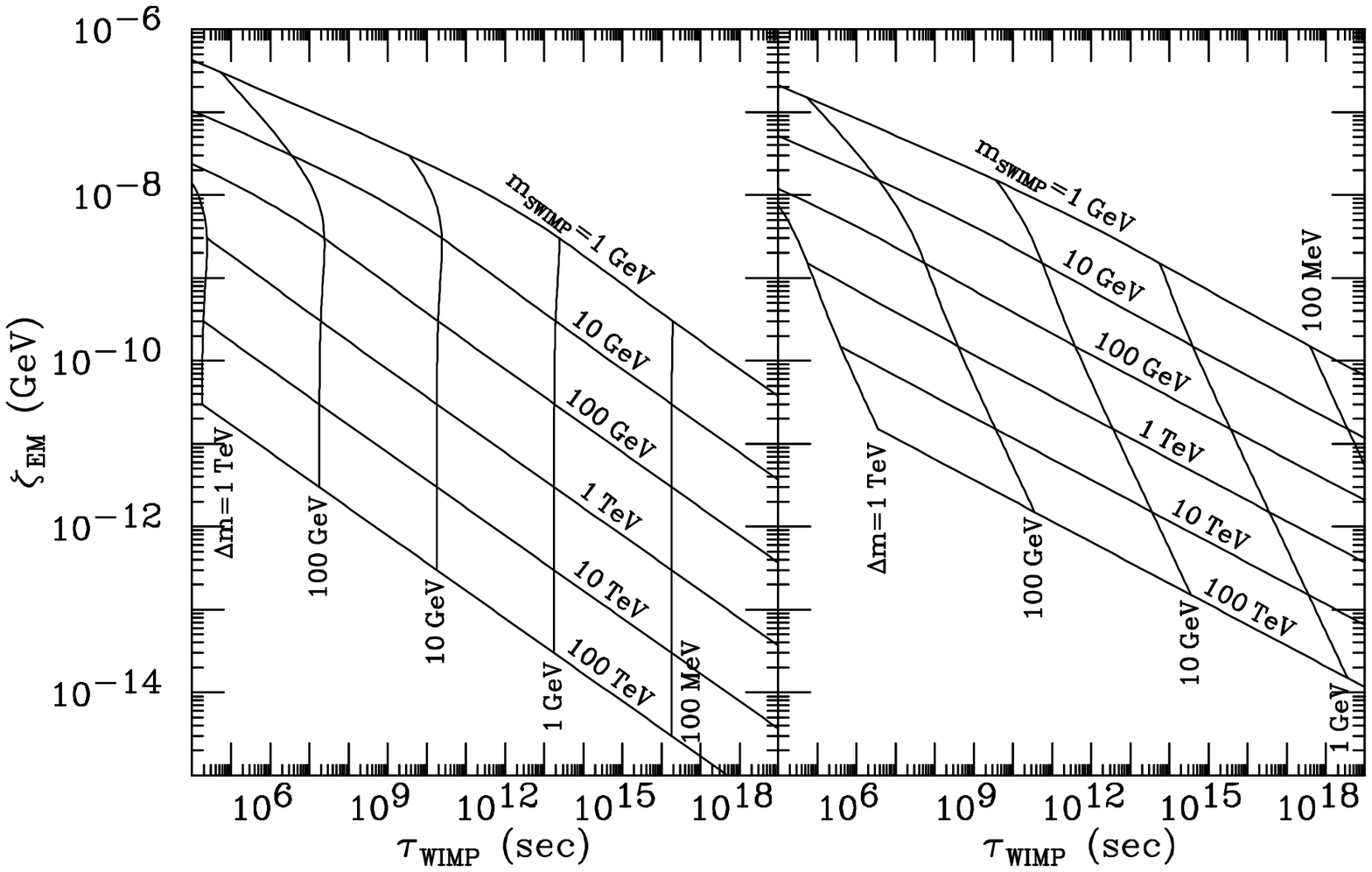}
\caption{Predicted values of WIMP lifetime $\tau$ and electromagnetic
energy release $\zetaEM \equiv \epsEM \YWIMP$ in the $\Bino$ (left)
and $\stau$ (right) NLSP scenarios for $\mSWIMP = 1~\gev$, $10~\gev$,
\ldots, $100~\tev$ (top to bottom) and $\Delta m \equiv \mWIMP -
\mSWIMP = 1~\tev$, $100~\gev$, \ldots, $100~\mev$ (left to right).
For the $\stau$ NLSP scenario, we assume $\epsEM = \frac{1}{2}
E_{\tau}$.  {}From Ref.~\cite{Feng:2003uy}.
\label{fig:prediction} }
\end{figure}

The energy release is conveniently expressed in terms of
\begin{eqnarray}
\xi_{\text{EM}} \equiv \epsilon_{\text{EM}} B_{\text{EM}}
Y_{\text{NLSP}}
\label{eq:xi_EM}
\end{eqnarray}
for electromagnetic energy, with a similar expression for hadronic
energy.  Here $\epsilon_{\text{EM}}$ is the initial EM energy released
in NLSP decay, and $B_{\text{EM}}$ is the branching fraction of NLSP
decay into EM components.  $Y_{\text{NLSP}} \equiv
n_{\text{NLSP}}/n_{\gamma}$ is the NLSP number density just before
NLSP decay, normalized to the background photon number density
$n_{\gamma} = 2 \zeta(3) T^3 / \pi^2$.  It can be expressed in terms
of the superWIMP abundance:
\begin{equation}
Y_{\text{NLSP}}\simeq 3.0 \times 10^{-12}
\left[\frac{\tev}{m_{\gravitino}}\right]
\left[\frac{\Omega_{\gravitino}}{0.23}\right] \ .
\label{eq:def_Y}
\end{equation}
Once an NLSP candidate is specified, and assuming superWIMPs make up
all of the dark matter, with $\Omega_{\gravitino} = \Omega_{\text{DM}}
= 0.23$, both the lifetime and energy release are determined by only
two parameters: $m_{\gravitino}$ and $m_{\NLSP}$.  The results for
slepton and neutralino NLSPs are given in \figref{prediction}.

\begin{figure}[tbp]
\includegraphics*[width=10cm]{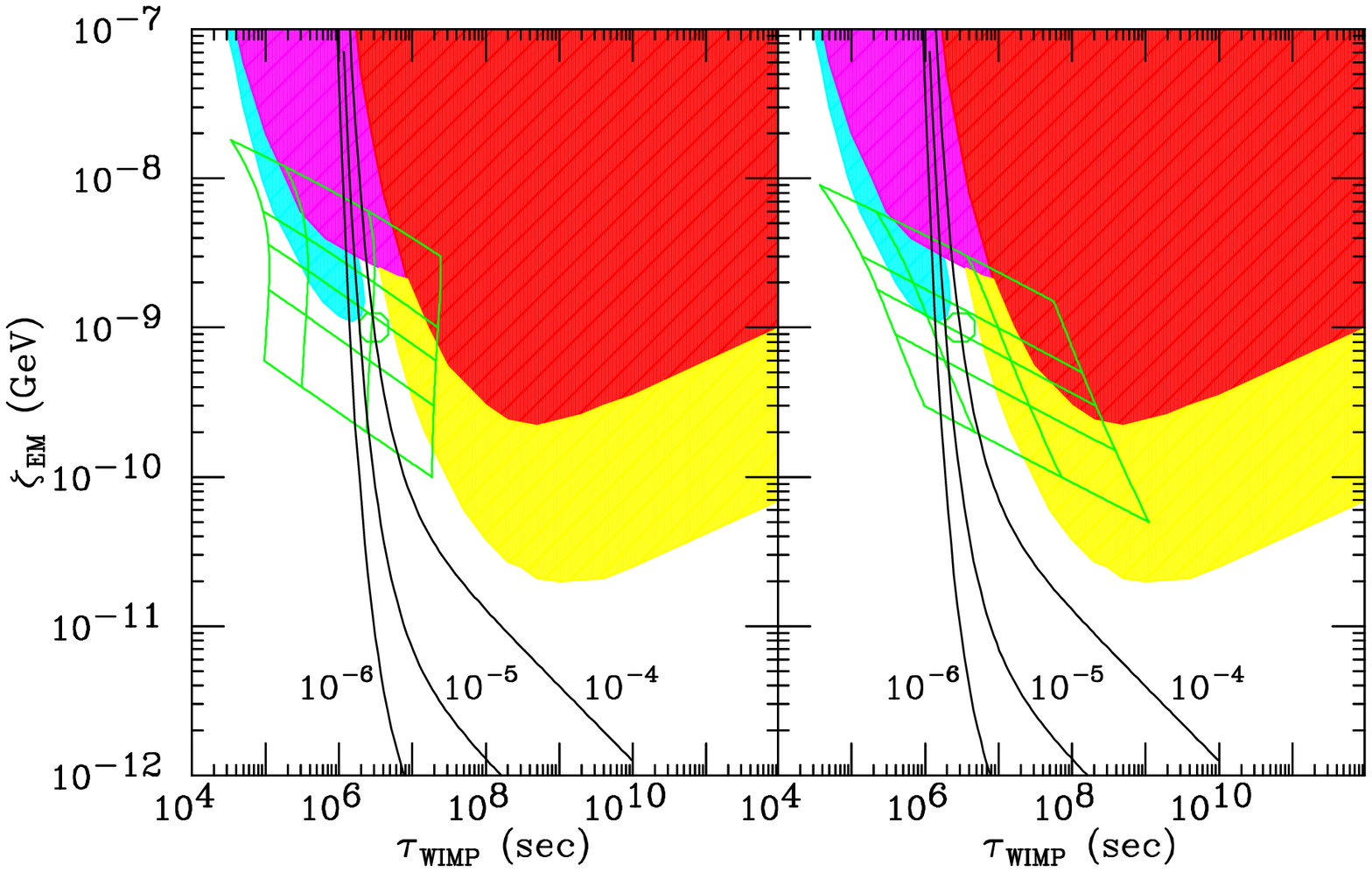}
\caption{The grid gives predicted values of WIMP lifetime $\tau$ and
electromagnetic energy release $\zetaEM \equiv \epsEM \YWIMP$ in the
$\Bino$ (left) and $\stau$ (right) WIMP scenarios for $\mSWIMP =
100~\gev$, $300~\gev$, $500~\gev$, $1~\tev$, and $3~\tev$ (top to
bottom) and $\Delta m \equiv \mWIMP - \mSWIMP = 600~\gev$, $400~\gev$,
$200~\gev$, and $100~\gev$ (left to right).  For the $\stau$ WIMP
scenario, we assume $\epsEM = \frac{1}{2} E_{\tau}$. The shaded
regions are excluded in one analysis of BBN
constraints~\protect\cite{Cyburt:2002uv}; the circle gives a region in
which $^7$Li is reduced to observed levels.  The contours are for
$\mu$, which parameterizes the distortion of the CMB from a Planckian
spectrum. {}From Ref.~\cite{Feng:2003uy}.
\label{fig:mu} }
\end{figure}

In \figref{mu}, these predictions are compared with BBN and CMB
constraints.  The shaded regions are excluded by an analysis of BBN
constraints on EM energy release~\protect\cite{Cyburt:2002uv}.  This
analysis has been strengthened by including hadronic constraints and
updated and refined in many ways in recent years, as described in
Chapter~28\rem{\ref{Chap:Jedamzik}}.  Although the excluded region has
shifted around, the basic features remain: some of the gravitino
superWIMP parameter space is excluded, and some remains.  In addition,
late decays to superWIMPs may in fact improve the current disagreement
of standard BBN predictions with the observed $^7$Li and $^6$Li
abundances~\cite{Cumberbatch:2007me,Bailly:2008yy}.

Figure \ref{fig:mu} also includes contours of the chemical potential
$\mu$, as determined by updating the analysis of Ref.~\cite{Hu:1993gc}.
The current bound is $\mu < 9\times
10^{-5}$~\cite{Fixsen:1996nj,Eidelman:2004wy}. Although
there are at present no indications of deviations from black body,
current limits are already sensitive to the superWIMP scenario, and
future improvements will further probe superWIMP parameter space.

\subsubsection{Small Scale Structure}

In contrast to WIMPs, superWIMPs are produced with large velocities at
late times.  This has two effects.  First, the velocity dispersion
reduces the phase space density, smoothing out cusps in DM halos.
Second, such particles damp the linear power spectrum, reducing power
on small
scales~\cite{Cembranos:2005us,Kaplinghat:2005sy,Borzumati:2008zz}.

\begin{figure}[tb]
\includegraphics[width=8cm]{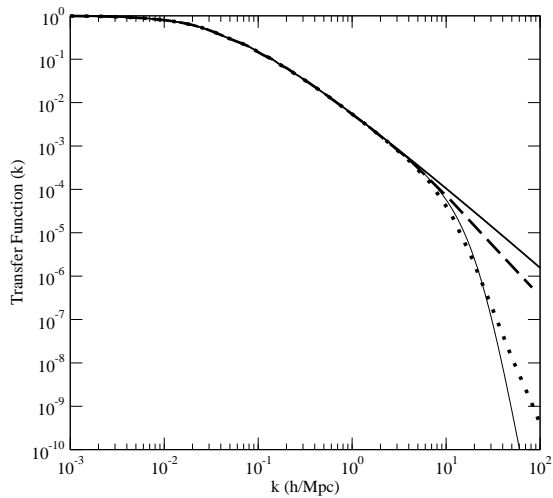}
\vspace*{-.2in}
\caption{The power spectrum for scenarios in which dark matter is
completely composed of WIMPs (solid), half WIMPs and half superWIMPs
(dashed), and completely composed of superWIMPs (dotted).  For
comparison, the this solid curve is the transfer function for a 1 keV
warm DM model.  {}From Ref.~\cite{Kaplinghat:2005sy}.
\label{fig:small_scale}
}
\end{figure}

Depending on the particular decay time and decay kinematics,
superWIMPs may be cold or warm.  As seen in \figref{small_scale},
superWIMPs may suppress small scale structure as effectively as a 1
keV sterile neutrino.  Some superWIMP scenarios may therefore been
differentiated from standard cold DM scenarios by studies of halo
profiles, and may even be favored by indications that cold DM predicts
halos that are too cuspy, as discussed in
Chapter~3\rem{\ref{Chap:Kaplinghat}}.

\subsection{Astroparticle and Collider Signals}

The possibility of long-lived charged particles in superWIMP scenarios
also has many implications for astroparticle and particle physics
experiments. 

\subsubsection{Cosmic Rays}

In superWIMP (and other similar) scenarios, long-lived charged
particles may be produced by cosmic rays, resulting in exotic signals
in cosmic ray and cosmic neutrino
experiments~\cite{Albuquerque:2003mi,Bi:2004ys,%
Albuquerque:2006am,Ahlers:2006pf,Ando:2007ds,Albuquerque:2008zs,%
Canadas:2008ey}.  As an example, ultra-high energy neutrinos may
produce events with two long-lived sleptons through $\nu q \to
\tilde{l} \tilde{q}'$ followed by the decay $\tilde{q}' \to
\tilde{l}$.  The sleptons are metastable and propagate to neutrino
telescopes~\cite{Huang:2006ie}, where they have a typical transverse
separation of hundreds of meters.  They may therefore be detected
above background as events with two upward-going, extremely high
energy charged tracks in experiments such as IceCube.

\subsubsection{Colliders}
\label{sec:colliders}

As evident in \eqref{decaylifetime}, in supersymmetric superWIMP
scenarios, the NLSP decays to the gravitino with lifetimes that may be
of the order of seconds to months.  Such particles are effectively
stable in collider experiments, and this scenario therefore implies
that each supersymmetric event will be characterized not by missing
energy, but by two charged, heavy metastable particles.  This is a
spectacular signal that will be cannot escape notice at the
LHC~\cite{Drees:1990yw,Goity:1993ih,%
Nisati:1997gb,Feng:1997zr,Feng:2005gj}.  In addition, given the
possibility of long lifetimes, it suggests that decays to gravitinos
may be observed by capturing slepton NLSPs and detecting their decays.

\begin{figure*}[tb]
\centering
\includegraphics[width=7cm]{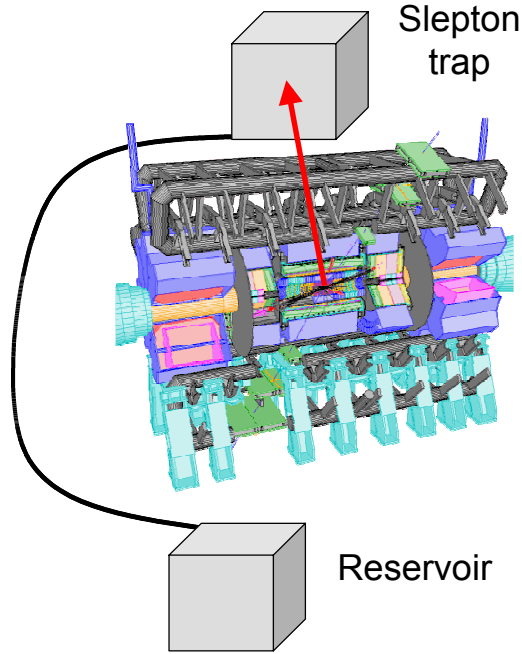}
\caption{Configuration for slepton trapping in gravitino superWIMP
scenarios.  {}From Ref.~\protect\cite{Feng:2004yi}.}
\label{fig:trap_diagram}
\end{figure*}

The sleptons may be captured in water tanks placed outside collider
detectors~\cite{Feng:2004yi}, in the detectors
themselves~\cite{Hamaguchi:2004df}, or by mining LHC cavern walls for
sleptons~\cite{DeRoeck:2005bw}.  In the first case, shown in
\figref{trap_diagram}, the water tanks may be drained periodically to
underground reservoirs where slepton decays may be observed in quiet
environments.  As many as $10^4$ sleptons per year may be stopped in 1
meter thick water tanks, opening up the possibility of a precise
measurement of slepton lifetime and the first study of a gravitational
process at high energy colliders, along with many other
implications~\cite{Feng:2004gn}.

\section{WIMPless Dark Matter}

\subsection{Candidates and Relic Densities}

Under general conditions, the thermal relic density of a particle $X$
is~\cite{Zeldovich:1965,Chiu:1966kg,Steigman:1979kw,Scherrer:1985zt}
\begin{equation}
\Omega_X \propto {1\over \langle \sigma v \rangle}
\sim \frac{m_X^2}{g_X^4} \ ,
\label{omega}
\end{equation}
where $\langle \sigma v \rangle$ is its thermally-averaged
annihilation cross section, and $m_X$ and $g_X$ are the characteristic
mass scale and coupling entering this cross section.  The last step
follows from dimensional analysis.  The WIMP miracle is the statement
that, for $m_X \sim \mweak \sim 100~\gev - 1~\tev$ and $g_X \sim
\gweak \simeq 0.65$, $\Omega_X$ is roughly $\Omega_{\text{DM}} \approx
0.23$.

Equation (\ref{omega}) makes clear, however, that the thermal relic
density fixes only one combination of the dark matter's mass and
coupling, and other combinations of $(m_X, g_X)$ can also give the
correct $\Omega_X$.  WIMPless models~\cite{Feng:2008ya} are those in
which the correct thermal relic density is achieved with parameters
$(m_X, g_X) \ne (\mweak, \gweak)$.

\begin{figure}[tb]
\begin{center}
\includegraphics*[width=10cm]{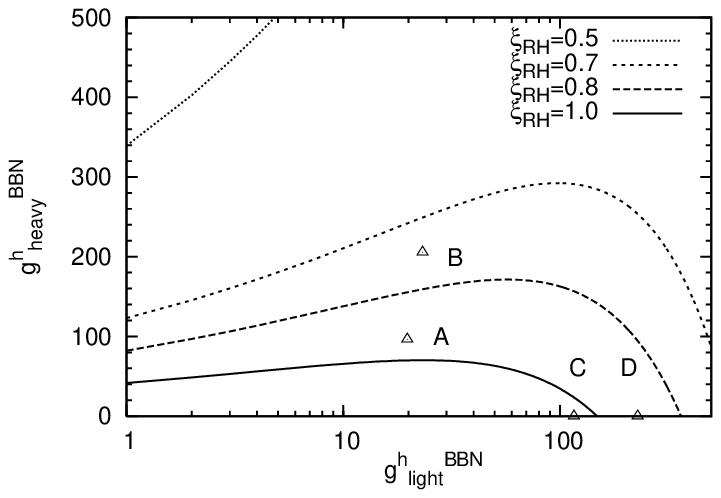}
\end{center}
\vspace*{-.29in}
\caption{Bounds from BBN in the $(\ghlightBBN, \ghheavyBBN)$ plane,
where $\ghlightBBN$ and $\ghheavyBBN$ are the hidden degrees of
freedom with masses $m < \ThBBN$ and $\ThBBN < m < \ThRH$,
respectively, for hidden to observable sector reheat temperature ratios
$\ThRH/\TRH = 0.5$, 0.7, 0.8, 1.0 (from top to bottom).  The regions
above the contours are excluded.  The values of $(\ghlightBBN,
\ghheavyBBN)$ are marked for four example hidden sectors: (A)
1-generation and (B) 3-generation flavor-free versions of the MSSM
with $\ThBBN < m_X < \ThRH$, and (C) 1-generation and (D) 3-generation
flavor-free versions of the MSSM with $m_X < \ThBBN/2$.  {}From
Ref.~\cite{Feng:2008mu}.  }
\label{fig:gstarlog}
\end{figure}

Because WIMPless dark matter does not have weak interactions, and
existing constraints effectively exclude electromagnetic and strong
interactions, WIMPless dark matter is necessarily hidden dark matter,
that is, dark matter that has no standard model gauge interactions.
Hidden sectors have a long history, and hidden sector dark matter has
been discussed for decades, beginning with work on mirror matter and
related ideas. For a general discussion and references, see
Chapter~9\rem{\ref{Chap:Servant}}.  Here we note only that, counter to
conventional wisdom, existing constraints place only weak bounds on
hidden sectors.  For example, light degrees of freedom change the
expansion rate of the Universe and thereby impact BBN.  The constraint
from BBN is highly sensitive to the temperature of the hidden sector,
however.  Current bounds from BBN on the number of light and heavy
degrees of freedom are given in \figref{gstarlog}.  For hidden sector
temperatures within a factor of 2 of the observable sector, hundreds
of degrees of freedom, equivalent to several copies of the standard
model or the minimal supersymmetric standard model (MSSM), may be
accommodated.

\begin{figure}[tb]
\begin{center}
\includegraphics*[width=9cm]{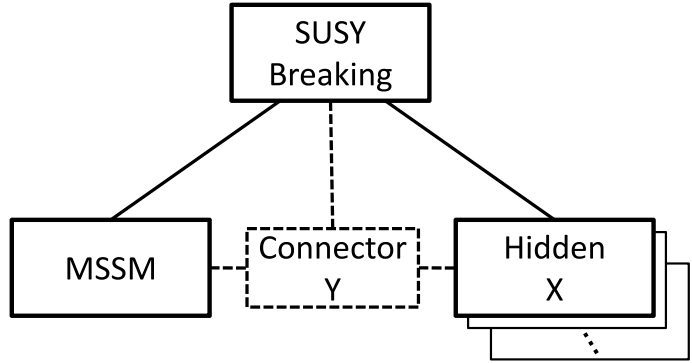}
\caption{Sectors of supersymmetric models.  Supersymmetry breaking is
mediated by gauge interactions to the MSSM and the hidden sector,
which contains the dark matter particle $X$.  An optional connector
sector contains fields $Y$, charged under both MSSM and hidden sector
gauge groups, which induce signals in direct and indirect searches and
at colliders.  There may also be other hidden sectors, leading to
multi-component dark matter.  {}From Ref.~\protect\cite{Feng:2008ya}.
}
\label{fig:sectors}
\end{center}
\end{figure}

Of course, WIMPless dark matter requires hidden sectors with
additional structure to guarantee that the hidden sector's dark matter
has the desired thermal relic density.  Remarkably, this structure may
be found in well-motivated models that have been explored previously
for many other reasons~\cite{Feng:2008ya}.  As an example, consider
supersymmetric models with gauge-mediated supersymmetry breaking
(GMSB)~\cite{Dine:1981za,Dimopoulos:1981au,Nappi:1982hm,%
AlvarezGaume:1981wy,Dine:1994vc,Dine:1995ag}.  These models
necessarily have several sectors, as shown in \figref{sectors}.  The
supersymmetry-breaking sector includes the fields that break
supersymmetry dynamically and mediate this breaking to other sectors
through gauge interactions.  The MSSM sector includes the fields of
the minimal supersymmetric standard model.  In addition, supersymmetry
breaking may also be mediated to one or more hidden sectors.  The
hidden sectors are not strictly necessary, but given the discussion
above, there is no reason to prevent them, and hidden sectors are
ubiquitous in such models originating in string theory.

As is well-known, GMSB models generate superpartner masses
proportional to gauge couplings squared.  Slightly more precisely, the
MSSM superpartner masses are
\begin{equation}
\label{mmass}
m \sim \frac{g^2}{16 \pi^2} \frac{F}{M} \ ,
\end{equation}
where $g$ is the largest relevant standard model gauge coupling, and
$F$ and $M$ are the vacuum expectation values of the
supersymmetry-breaking sector's chiral field $S$, with $\langle S
\rangle = M + \theta^2 F$.  With analogous couplings of the hidden
sector fields to hidden messengers, the hidden sector superpartner
masses are
\begin{equation}
\label{mxmass}
m_X \sim \frac{g_X^2}{16 \pi^2} \frac{F}{M} \ ,
\end{equation}
where $g_X$ is the relevant hidden sector gauge coupling.  As a
result,
\begin{equation}
\frac{m_X}{g_X^2} \sim \frac{m}{g^2} \sim
\frac{F}{16 \pi^2 M} \ ;
\label{mxgx}
\end{equation}
that is, $m_X/g_X^2$ is determined solely by the
supersymmetry-breaking sector.  As this is exactly the combination of
parameters that determines the thermal relic density of \eqref{omega},
the hidden sector automatically includes a dark matter candidate that
has the desired thermal relic density, irrespective of its mass.

\begin{figure}[tb]
\begin{center}
\includegraphics*[width=10cm]{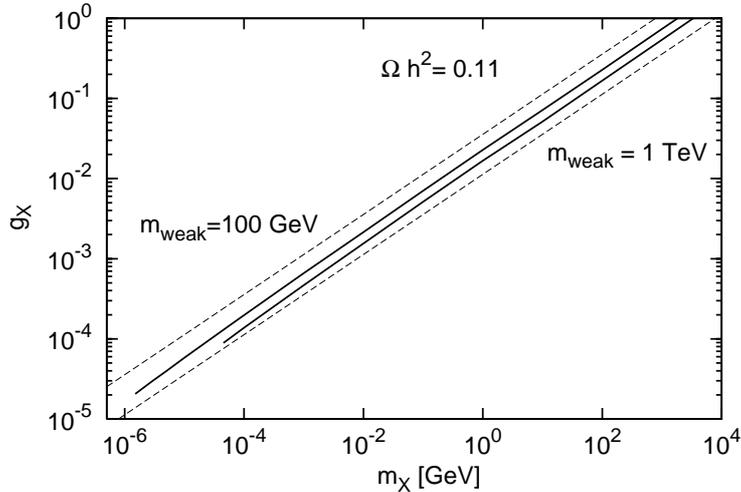}
\end{center}
\vspace*{-.29in}
\caption{Contours of $\Omega_X h^2 = 0.11$ in the $(m_X,g_X)$ plane
for hidden to observable reheat temperature ratios $\ThRH/\TRH=0.8$
(upper solid) and 0.3 (lower solid), where the hidden sector is a
1-generation flavor-free version of the MSSM.  Also plotted are lines
of $\mweak \equiv (m_X/g^2_X)g'^2 = 100~\gev$ (upper dashed) and 1 TeV
(lower dashed).  {}From Ref.~\cite{Feng:2008mu}.  }
\label{fig:gxmx}
\end{figure}

The freeze out of hidden sector dark matter in such GMSB models has
been studied numerically.  As an example, in Ref.~\cite{Feng:2008mu}
the hidden sector was assumed to be a copy of the MSSM, but with a
free superpartner mass scale $m_X$ and all Yukawa couplings $\sim
{\cal O}(1)$, so that the only light hidden sector particles are the
hidden gluon, photon, and neutrinos.  The results are given in
\figref{gxmx}.  The criterion that the standard model weak scale be
between 100 GeV and 1 TeV requires values of $(m_X, g_X)$ within the
band.  The solid curves, where the thermal relic density of hidden
dark matter is consistent with dark matter, are seen to lie within this
band, confirming the scaling arguments and rough estimates described
above.

In summary, well-known frameworks for hidden sectors include models in
which the hidden sector contains a particle whose thermal relic
density is automatically in the desired range to be dark matter, even
when the particle's mass is not at the weak scale.  This property
relies on the relation $m_X \propto g_X^2$, which is common to other
frameworks for new physics that avoid flavor-changing problems, such
as anomaly-mediated supersymmetry breaking.  The ``coincidence''
required for WIMPless dark matter may also be found in other settings;
see, for example, Ref.~\cite{Hooper:2008im}.

\subsection{Direct and Indirect Detection Signals}

The decoupling of the WIMP miracle from WIMPs has many possible
implications and observable consequences.  In the case that the dark
matter is truly hidden, it implies that there are no prospects for
direct or indirect detection.  Signals must be found in astrophysical
observations, as in the case of superWIMPs.  Alternatively, there may
be connector sectors containing particles that mediate interactions
between the standard model and the hidden sector through non-gauge
(Yukawa) interactions (see \figref{sectors}).  Such connectors may
generate many signals with energies and rates typically unavailable to
WIMPs.

As an example, first consider direct detection.  The DAMA signal,
interpreted as spin-independent, elastic scattering, has
conventionally favored a region in the mass-cross section plane with
$(m_X, \sigmaSI) \sim (20 - 200~\gev,
10^{-5}~\pb)$~\cite{Belli:1999nz}.  This is now excluded, most
stringently by XENON10~\cite{Angle:2007uj} and CDMS
(Ge)~\cite{Ahmed:2008eu}, which require $\sigmaSI < 10^{-7}~\pb$
throughout this range of $m_X$.  Gondolo and Gelmini have noted,
however, that an alternative region with $(m_X, \sigmaSI) \sim
(1-10~\gev, 10^{-3}~\pb)$ may explain the DAMA results without
violating other known bounds~\cite{Gondolo:2005hh}. DAMA's relative
sensitivity to this region follows from its low energy threshold and
the lightness of Na nuclei.  This region may be extended to lower
masses and cross sections by the effects of
channeling~\cite{Bernabei:2007hw,Petriello:2008jj,Bottino:2008mf,%
Savage:2008er,Fairbairn:2008gz} and may also be broadened by dark
matter streams in the solar neighborhood~\cite{Gondolo:2005hh}.

\begin{figure}[tb]
\includegraphics*[width=10cm]{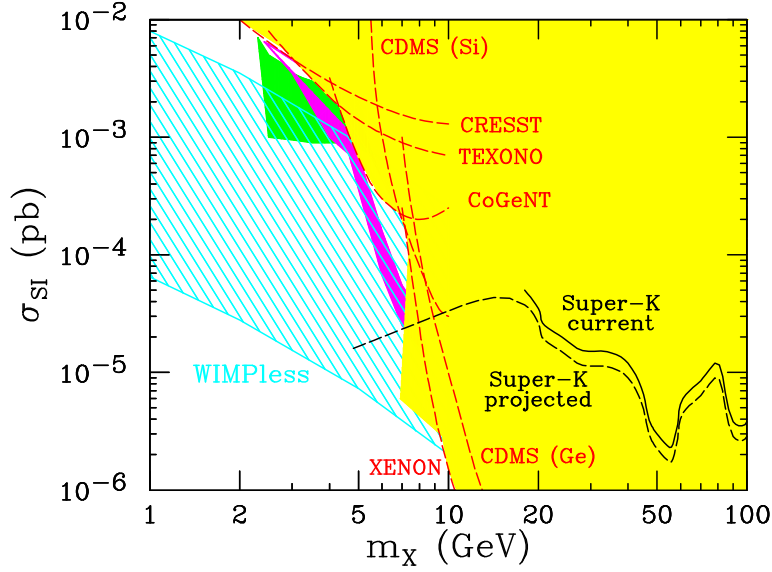}
\caption{Direct detection cross sections for spin-independent
$X$-nucleon scattering as a function of dark matter mass $m_X$.  The
magenta shaded region is DAMA-favored given channeling and no
streams~\cite{Petriello:2008jj}, and the medium green shaded region is
DAMA-favored at 3$\sigma$ given streams and no
channeling~\cite{Gondolo:2005hh}.  The light yellow shaded region is
excluded by the direct detection experiments indicated.  The blue
cross-hatched region is the parameter space of WIMPless models with
connector mass $m_Y = 400~\gev$ and $0.3 <\lambda_b < 1.0$.  The black
solid line is the published Super-K exclusion
limit~\cite{Desai:2004pq}, and the black dashed line is a projection
of future Super-K sensitivity.  {}From Ref.~\cite{Feng:2008qn}.
\label{fig:superkdirect2_mod}
}
\end{figure}

The acceptable DAMA-favored region with $m_X \sim 5~\gev$ has masses
that are low for WIMPs, given that their masses are expected to be
around $\mweak \sim 100~\gev - 1~\tev$.  However, in WIMPless models,
where the thermal relic density is achieved for a variety of dark
matter masses, such masses are perfectly natural.  A WIMPless particle
$X$ may couple to the standard model through Yukawa interactions
\begin{equation}
{\cal L} = \lambda_f X \bar{Y}_L f_L
+ \lambda_f X \bar{Y}_R f_R \ ,
\label{connector}
\end{equation}
where $Y$ is a vector-like connector fermion, and $f$ is a standard
model fermion.  Taking $f$ to be the $b$ quark, and the $Y$ mass to be
400 GeV, consistent with current bounds, these couplings generate
spin-independent scattering cross sections given in
\figref{superkdirect2_mod}.  We see that WIMPless dark matter may
explain the DAMA results without difficulty.

WIMPless dark matter also provides new target signals for indirect
detection.  For WIMPs, annihilation cross sections determine both the
thermal relic density and indirect detection signals.  The thermal
relic density therefore constrains the rates of indirect detection
signals.  In the WIMPless case, however, this connection is weakened,
since the thermal relic density is governed by hidden sector
annihilation and gauge interactions, while the indirect detection
signals are governed by the interactions of \eqref{connector}.

This provides a wealth of new opportunities for indirect detection.
As an example, WIMPless dark matter may be detected through its
annihilation to neutrinos in the Sun by experiments such as
Super-Kamiokande.  Although such rates depend on the competing cross
sections for capture and annihilation, the Sun has almost certainly
reached its equilibrium state, and the annihilation rate is determined
by the scattering cross section~\cite{Desai:2004pq}.  The prospects
for Super-Kamiokande may therefore be compared to direct detection
rates~\cite{Desai:2004pq,Hooper:2008cf,Feng:2008qn}.  The results are
given in \figref{superkdirect2_mod}.  In the near future,
Super-Kamiokande may be able to probe the low mass regions
corresponding to the DAMA signal.

WIMPless dark matter also provides additional targets for indirect
detection experiments looking for photons, positrons, and other
annihilation products.  The connectors may also play an interesting
role in collider experiments.  Further details may be found in
Refs.~\cite{Feng:2008dz,Hooper:2008cf,Feng:2008qn}.



\end{document}